\documentclass[fleqn,10pt]{wlscirep}
\usepackage[utf8]{inputenc}
\usepackage[T1]{fontenc}
\usepackage{placeins}
\usepackage{setspace}
\usepackage{lineno}

\title{The Logarithmic Memristor-Based Bayesian Machine}

\author[1]{Cl\'ement~Turck}
\author[1]{Kamel-Eddine~Harabi}
\author[1]{Adrien~Pontlevy}
\author[1]{Th\'eo~Ballet}
\author[2]{Tifenn~Hirtzlin}
\author[2,]{Elisa~Vianello} 
\author[3]{Rapha\"el~Laurent}
\author[3,4]{Jacques~Droulez}
\author[4]{Pierre~Bessi\`ere}
\author[5]{Marc~Bocquet}
\author[5]{Jean-Michel~Portal}
\author[1,*]{Damien~Querlioz}
\affil[1]{Universit\'e Paris-Saclay, CNRS, Centre de Nanosciences et de Nanotechnologies, 91120 Palaiseau, France}
\affil[2]{CEA, LETI, Universit\'e Grenoble-Alpes, 38400 Grenoble, France}
\affil[3]{HawAI.tech, 38000 Grenoble, France}
\affil[4]{Institut des Syst\`emes Intelligents et de Robotique, Sorbonne Universit\'e, CNRS, 75005 Paris, France}
\affil[5]{Aix Marseille Univ, CNRS, IM2NP, Marseille, France}
\affil[*]{damien.querlioz@universite-paris-saclay.fr}

\begin{abstract}
The  demand for  explainable and energy-efficient artificial intelligence (AI) systems for edge computing has led to significant interest in electronic systems dedicated to Bayesian inference. Traditional designs of such systems often rely on stochastic computing, which offers high energy efficiency but suffers from latency issues and struggles with low-probability values. In this paper, we introduce the logarithmic memristor-based Bayesian machine, an innovative design that leverages the unique properties of memristors and logarithmic computing as an alternative to stochastic computing. We present a prototype machine fabricated in a hybrid CMOS/hafnium-oxide memristor process. We validate the versatility and robustness of our system through  experimental validation and extensive simulations in two distinct applications: gesture recognition and sleep stage classification. The logarithmic approach simplifies the computational model by converting multiplications into additions and enhances the handling of low-probability events, which are crucial in time-dependent tasks. Our results demonstrate that the logarithmic Bayesian machine achieves superior performance in terms of accuracy and energy efficiency compared to its stochastic counterpart, particularly in scenarios involving complex probabilistic models. This work paves the way for the deployment of advanced AI capabilities in edge devices, where power efficiency and reliability are paramount.
\end{abstract}
\begin{document}
\maketitle

%
\thispagestyle{empty}


\section*{Introduction}
 
The rapid evolution of artificial intelligence (AI) applications at the edge has accentuated the demand for low-power, explainable, and reliable edge AI systems that can function effectively even in uncertain conditions \cite{rai2020explainable}. Numerous works have shown that novel memory devices allow for considerable energy savings for the implementation of neural networks, a non-explainable form of AI, through the cointegration of logic and memory \cite{ielmini2018memory,spiga2020memristive,ambrogio2023analog,aguirre2024hardware,huang2024memristor}.
This cointegration can also be applied to explainable AI, as exemplified by the recently demonstrated nanodevice-based Bayesian machines  \cite{harabi2023memristor,gong2023first}. Relying on Bayesian inference \cite{jaynes2003probability,bessiere2013bayesian}, the strength of these machines are manifold: they provide a hardware AI with inherently explainable results \cite{ghahramani2015probabilistic,letham2015interpretable}, at a high energy efficiency.
Bayesian machines are particularly appealing for tasks where neural networks struggle, such as sensor fusion in highly uncertain environments with limited training data, or safety-critical applications requiring explainable decisions.

The efficiency of Bayesian machines is achieved through near-memory computing, which offsets the high cost of accessing the parameters of the Bayesian model \cite{pedram2017dark}, and the adoption of stochastic computing \cite{gaines1969stochastic,alaghi2013survey}. The latter particularly facilitates compact and energy-efficient multiplication, a dominant arithmetic operation in the Bayesian machine and has been shown repeatedly to be particularly adapted to Bayesian inference \cite{friedman2016bayesian,fernandes2016bayesian,faix2016design,shim2017stochastic,faria2018implementing,jia2019spinbis,hoe2019bayesian,debashis2020hardware,zheng2022hardware}.
  Despite all their merits, Bayesian machines are not exempt from the challenges of stochastic computing. These include high latency and compromised precision when dealing with low probability values \cite{alaghi2013survey,sousa2021nonconventional}. While Bayesian machines demonstrated high accuracy in a gesture recognition application \cite{harabi2023memristor} or on handwritten character recognition \cite{gong2023first},  their broader applicability to diverse  tasks remains an open question.

Our contribution through this work is two-fold. First, we introduce an alternative design: the  logarithmic memristor-based Bayesian machine. This design eschews stochastic computing in favor of logarithmic computing \cite{sousa2021nonconventional}. This modification translates to implementing probability multiplications using digital near-memory integer adders, thereby augmenting the versatility of the system. We present a fully fabricated logarithmic Bayesian machine, using a hybrid CMOS/hafnium-oxide memristor process and show its robustness experimentally. Second, we undertake a comprehensive evaluation of this machine,  for gesture recognition, and also for a Bayesian filter application that addresses a time-dependent task: sleep stage classification throughout the night. A detailed comparison with the stochastic design, assessing accuracy and energy consumption,  elucidates the scenarios in which each design excels.

Preliminary measurements of the logarithmic Bayesian machine were presented at recent conferences \cite{turck2023energy,turck2023bayesian}. This paper, based on new measurements and simulations, adds extensive comparisons between the logarithmic and the stochastic versions of the Bayesian machines.


\section*{Results}

\subsection*{Bayesian machines: the logarithmic approach}

\begin{figure}[h!]
\centering
\includegraphics[width=0.95\linewidth]{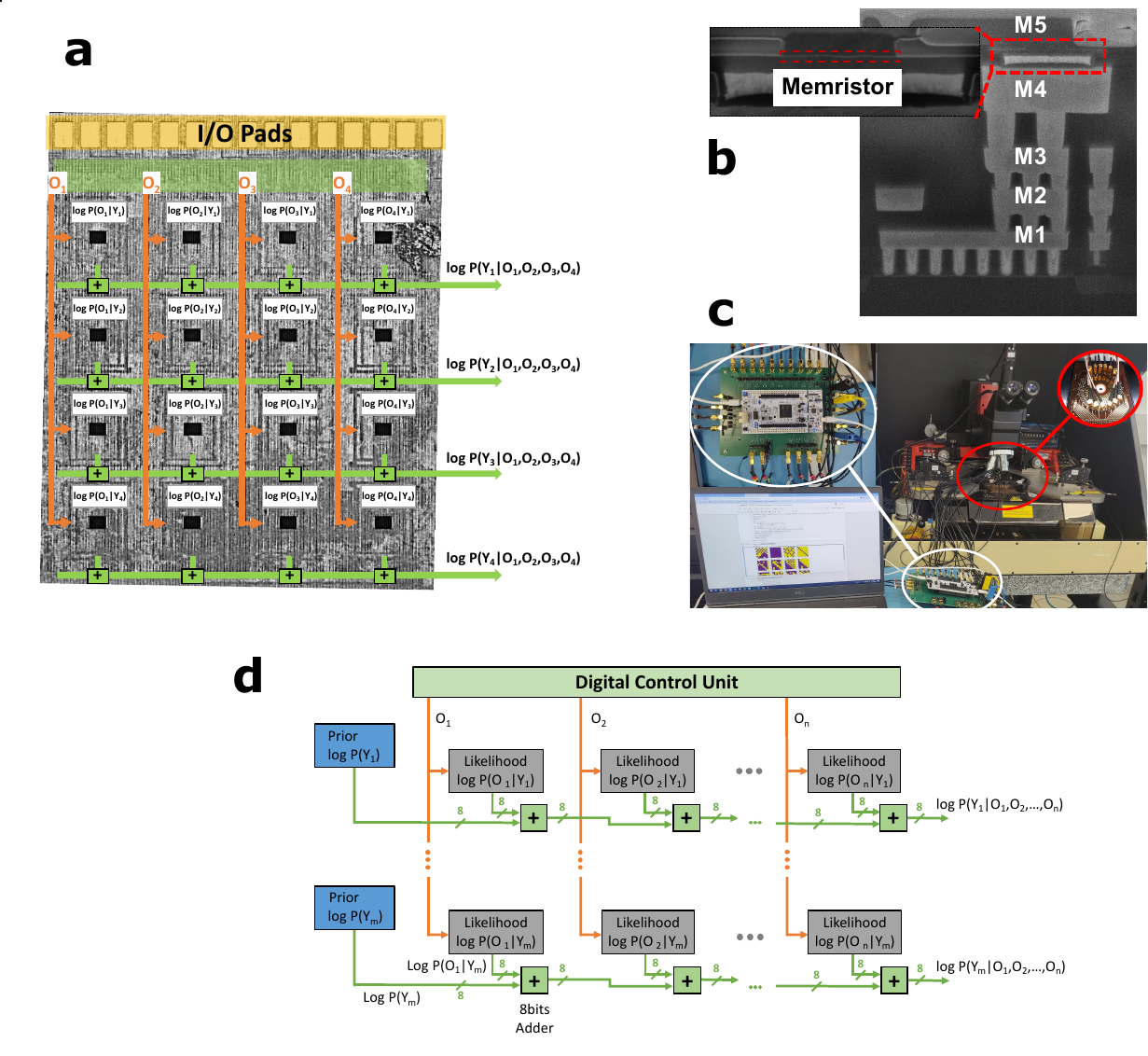}
\caption{
\textbf{General architecture of the logarithmic Bayesian machine.}
\textbf{a} Optical microscopy image of the die of the fabricated machine.
\textbf{b} Electron microscopy image of a memristor integrated in the CMOS backend of line of our process.
\textbf{c} Photograph of the logarithmic Bayesian machine test setup.
\textbf{d} Simplified schematic of the implemented logarithmic Bayesian machine.
All log-probabilities are coded as eight-bit integers, following eq.~\ref{eq:entier}.
}
\label{fig:general_archi}
\end{figure}

The primary objective of a Bayesian machine is performing the inference of Bayesian models, specifically Bayesian networks. Its versatile nature allows adaptation to various model types. One of the more straightforward implementations of this is the time-independent naive Bayesian inference, as highlighted in ref.~\cite{harabi2023memristor}. We aim at inferring the value $y$ of a variable $Y$ based on a collection of observations $O_1$, $O_2$,..., $O_n$. The probability distribution of $Y$ is given by
%
\begin{equation}
p(Y=y \mid O_1, O_2,\ldots,O_n) \propto  p(O_1 \mid Y=y) \times p(O_2\mid Y=y) \times  p(O_3\mid Y=y) \times  \ldots  \times  p(O_n\mid Y=y)  \times p(Y=y).
\label{eq:naivebayeslaw}
\end{equation}
The likelihoods $p(O_i\mid Y=y)$,  as well as the priors $p(Y=y)$ constitute the parameters of the model.

In the stochastic Bayesian machine, stochastic computing is employed to compute the product of probabilities in equations such as eq.~\ref{eq:naivebayeslaw}. All probability values are represented by a stochastic bit stream whose probability of being one is the coded probability. Then, multiplication is achieved with great energy efficiency and compactness by utilizing only AND gates \cite{gaines1969stochastic,alaghi2013survey,harabi2023memristor}. 

The Bayesian machine is adaptable to more complex, non-naive models, as detailed in the first Supplementary Note of ref. \cite{harabi2023memristor}. However, the stochastic machine possesses inherent challenges. First, it needs random number generators (RNGs). While these RNGs can be shared column-wise to mitigate their cost, energy is spent for transmitting the random numbers throughout the column. Second, when likelihoods are small, stochastic computing can require an unreasonably high number of clock cycles to provide accurate results \cite{alaghi2013survey}. 

In this work, we explore an alternative approach, which transforms products into sums. All probability values are represented logarithmically using integers. The correspondence between a real probability $p$ and the integer $n$ is:
\begin{equation}
  p \approx B^{\frac{n}{m}}.
\label{eq:entier}
\end{equation}

Throughout our research, we use 8-bit integers to represent $n$, and we selected $B=1/2$ and $m=8$. This encoding implies that integer values from 0 to 8 correspond to probabilities ranging between 1 and 1/2, while values from 9 to 255 represent probabilities smaller than 1/2. The smallest probability that can be coded is $\frac{1}{2}^{255/8}\approx2.5\times10^{-10}$. Therefore, this encoding prioritizes lower probability values, contrasting with the methodology of stochastic computing, where such values are difficult to process.

\subsection*{Sleep stage classification using a fabricated hybrid memristor/CMOS logarithmic Bayesian Machine}

Our logarithmic Bayesian machine, as depicted in Fig.~\ref{fig:general_archi}d, retains the architectural principle of its stochastic counterpart. Memristor arrays host logarithmic integer representations of the likelihoods  $p(O_i \mid Y=y)$, with storage as 8-bit integers. The input observations $O_i$ provide row addresses for these arrays, indicating the specific likelihood value to access. These values are then aggregated using near-memory 8-bit integer adders, which substitute the 1-bit AND gates found in the stochastic model. Our adders are designed to yield a 255 output (i.e., the minimum possible probability) in overflow scenarios.

Employing a hybrid CMOS/memristor process, we fabricated a logarithmic Bayesian machine test chip. The memristors, also referred to as resistive RAM, ReRAM, or RRAM in industrial contexts, comprise a stack of titanium nitride, hafnium oxide,  titanium, and titanium nitride. A detailed overview of the process is presented in the Methods section. Fig.~\ref{fig:general_archi}a displays the optical image of the die, which includes 16 memristor array blocks. Its layout closely aligns with the schematic of Fig.~\ref{fig:general_archi}d. The memristors are embedded in the backend of line of the CMOS, positioned between metal levels four and five, as illustrated in the electron microscopy image of Fig.~\ref{fig:general_archi}b.

Despite their functionality, memristors are not flawless. Their imperfections can lead to bit errors, which are eliminated using strong error correction codes (ECC) in industrial designs \cite{chang2020envm,wu2023emerging}. In our design, we use an alternative approach. To counteract errors emerging from memristor variability \cite{ly2018role} and instability \cite{esmanhotto2022experimental}, we have used a two-transistor / two-memristor (2T2R) bit cell configuration. In this cell, bits are programmed complementarily. This technique, adopted previously in the stochastic Bayesian machine \cite{harabi2023memristor} and in a memristor-based binarized neural network \cite{jebali2024powering}, has demonstrated a significant reduction in bit error rate, rivalling the efficiency of a single-error-correction, double-error-detection (SECDED) ECC with equivalent redundancy \cite{hirtzlin2020digital}. The 2T2R cells are read using precharge sense amplifiers \cite{zhao2009high,zhao2014synchronous}. Each column of the memristor array features its dedicated sense amplifier. 

The test chip also houses comprehensive control circuitry for programming the memristors (detailed in Methods). Furthermore, level shifters located near each memory array enable the application of high voltages essential for memristor forming and programming, all managed through nominal voltage control signals (as described in Methods). We have thoroughly probe tested the circuit, aided by a specialized printed circuit board that facilitates comprehensive computer control, presented in Fig.~\ref{fig:general_archi}c (see Methods).

\begin{figure}[h!]
\centering
\includegraphics[width=0.7\linewidth]{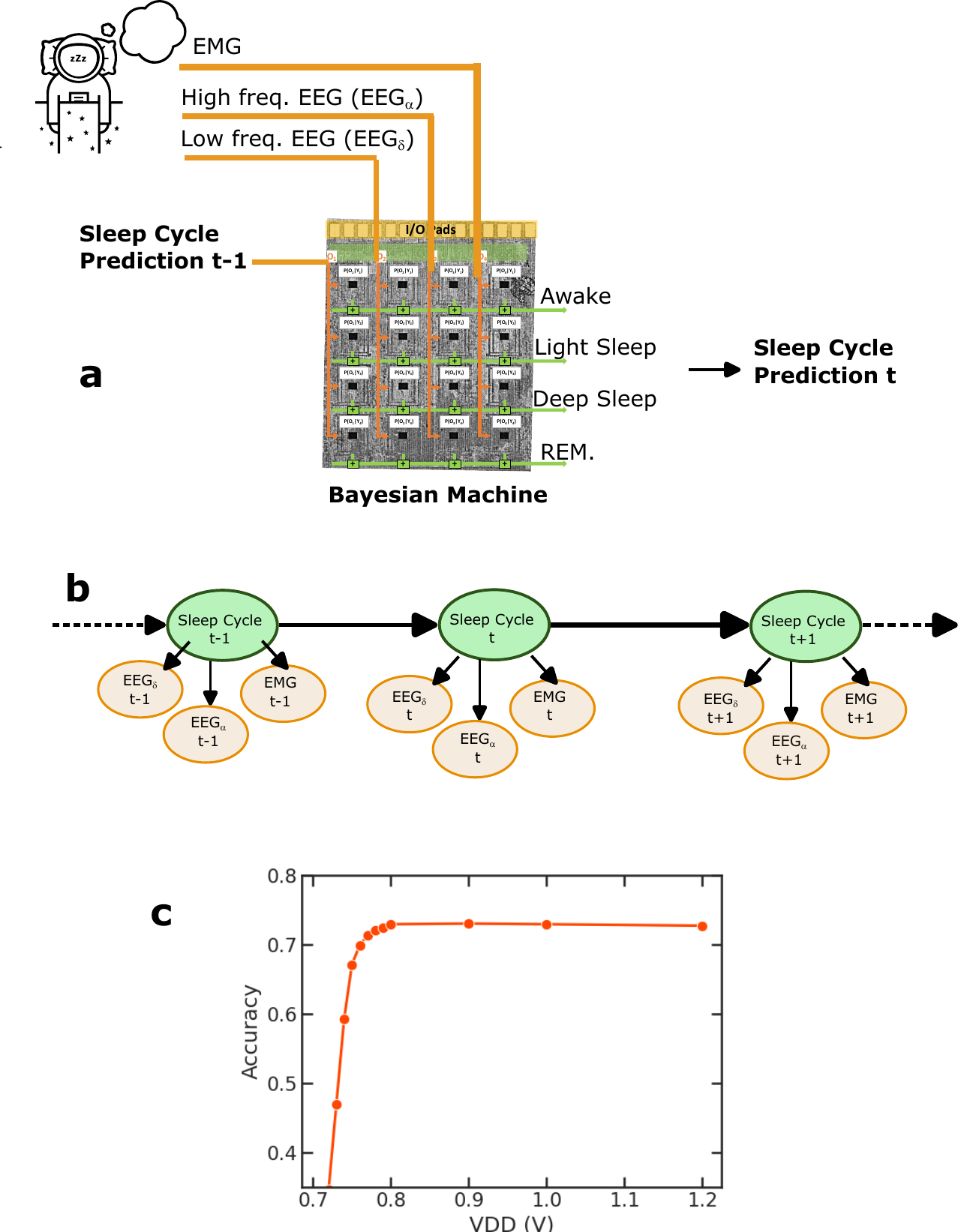}
\caption{
\textbf{Experimental implementation of sleep stage classification throughout the night.}
\textbf{a} Inputs and and outputs of the logarithmic Bayesian machine used for sleep stage classification.
\textbf{b} Bayesian network model implemented on the logarithmic Bayesian machine used for sleep stage classification.
\textbf{c} Experimentally measured test accuracy of the logarithmic Bayesian machine, averaged over 1000 five-second segments. The measurement was repeated for various supply voltages VDD.
}
\label{fig:sleep_stage}
\end{figure}

In this study, we have employed sleep stage detection throughout the night as a benchmark task to evaluate our integrated circuit. Mathematically, this task is more intricate than the gesture recognition task used for evaluating the stochastic Bayesian machine in ref.~\cite{harabi2023memristor}. A typical night's sleep comprises different stages: awake, Rapid Eye Movement (REM), light sleep, and deep sleep. Note that for simplicity, we have aggregated  ``light'' and ``deep'' sleep stages, which are sometimes each separated into two separate stages (refer to Methods for details). Our objective is to classify sleep stages throughout the night using electroencephalography (EEG) and electromyography (EMG) measurements.

Inferring sleep stages solely from EEG and EMG readings is challenging. The temporal context plays a significant role: if a patient is identified to be in deep sleep at time $t$, the likelihood of the same stage five seconds later (denoted as $t+1$) is substantial. This sequential correlation is captured using a Bayesian model, as represented by the following relation, obtained using Bayes' law:
\begin{equation}
p\left(Y(t) = y \mid \text{EEG}, \text{EMG}, Y(t-1)\right) 
\propto p\left(\text{EEG}, \text{EMG} \mid Y(t) = y, Y(t-1)\right) \times p\left(Y(t) = y \mid Y(t-1)\right).
\label{eq:sleep1}
\end{equation}

Building on the assumption that both EEG and EMG data are primarily influenced by the sleep state at time $t$, and that this influence remains constant, we refine the equation as  
\begin{equation}
p\left(Y(t) = y \mid Y(t-1), \text{EEG}, \text{EMG}\right) 
\propto p\left(Y(t) = y \mid Y(t-1) \right) \times p\left(\text{EEG}, \text{EMG} \mid Y(t) = y\right).  
\label{eq:sleep2}
\end{equation}

This temporal Bayesian model is an instance of a ``Bayesian filter'' \cite{bessiere2013bayesian}. For our experiments, we extracted three key observations: the full power of the EMG signal, the EEG signal's power at 1.5~Hz (indicative of delta brain waves), and the EEG signal's power at 9.35~Hz (indicative of alpha brain waves). Detailed methodologies for these extractions, along with their implications, are provided in the Methods section. We used patient data from the DREAMS dataset \cite{devuyst2005dreams}, excluding the electrooculography data available therein. Our implementation offers a proof-of-concept case study for our Bayesian machine, which, while effective, does not maximize the potential accuracy achievable through a more comprehensive use of the DREAMS data. Assuming conditional independence among our three observations, our refined Bayesian model is then represented as:

\begin{align}
p\left(Y(t) = y \mid Y(t-1), \text{EEG}, \text{EMG}\right) 
\propto & \, p\left(Y(t) = y \mid Y(t-1)\right) \nonumber \\
&\times p\left(\text{EEG}_\delta \mid Y(t) = y\right) \nonumber \\
&\times p\left(\text{EEG}_\alpha \mid Y(t) = y\right) \nonumber \\
&\times p\left(\text{EMG} \mid Y(t) = y\right). 
\label{eq:sleep2}
\end{align}

More formally, this model performs inference over the Bayesian network presented in Fig.~\ref{fig:sleep_stage}b. 
Implementing this equation in our Bayesian machine is intuitive, with the preceding Bayesian prediction serving as an input and the three observations as the three other inputs of the machine (see Fig.~\ref{fig:sleep_stage}a). 
It is worth noting that the preceding Bayesian prediction has only four possible values, plus one unknown or uncertain prediction case used for the first prediction. Therefore, the memory arrays in the first column of our machine, which each possess eight rows, are programmed only partially.

We trained our Bayesian model on the DREAMS dataset, converted it to its logarithmic form, and subsequently loaded it into our Bayesian machine for experimental validation (see Methods for the details on each of these operations). We assessed the performance of our machine using the last 1000 measurements of a patient's night from the DREAMS dataset (which were not used for training the machine). The subset of 1000 measurements was chosen to strike a balance between comprehensiveness and experimental feasibility.

Our findings, presented in Fig.~\ref{fig:sleep_stage}c, reveal a 72\% accuracy at the chip's nominal supply voltage of 1.2~volts, aligning with software simulations of the machine. Remarkably, this accuracy was sustained even when the supply voltage was reduced to 0.8~volts, without the need for any chip recalibration. This robust performance can be attributed to the differential nature of our precharge sense amplifier, ensuring resilient memristor readings as thoroughly examined in ref.~\cite{jebali2024powering}. However, at even lower supply voltages, accuracy begins to decline due to read errors in the memristors. The prevalence of these errors at reduced voltages can be traced back to the high threshold voltage inherent to the low-power CMOS process we employed \cite{harabi2023memristor,jebali2024powering}. Therefore, even lower-supply voltage is foreseeable using a lower threshold voltage CMOS process.

\begin{figure}[h!]
\centering
\includegraphics[width=0.95\linewidth]{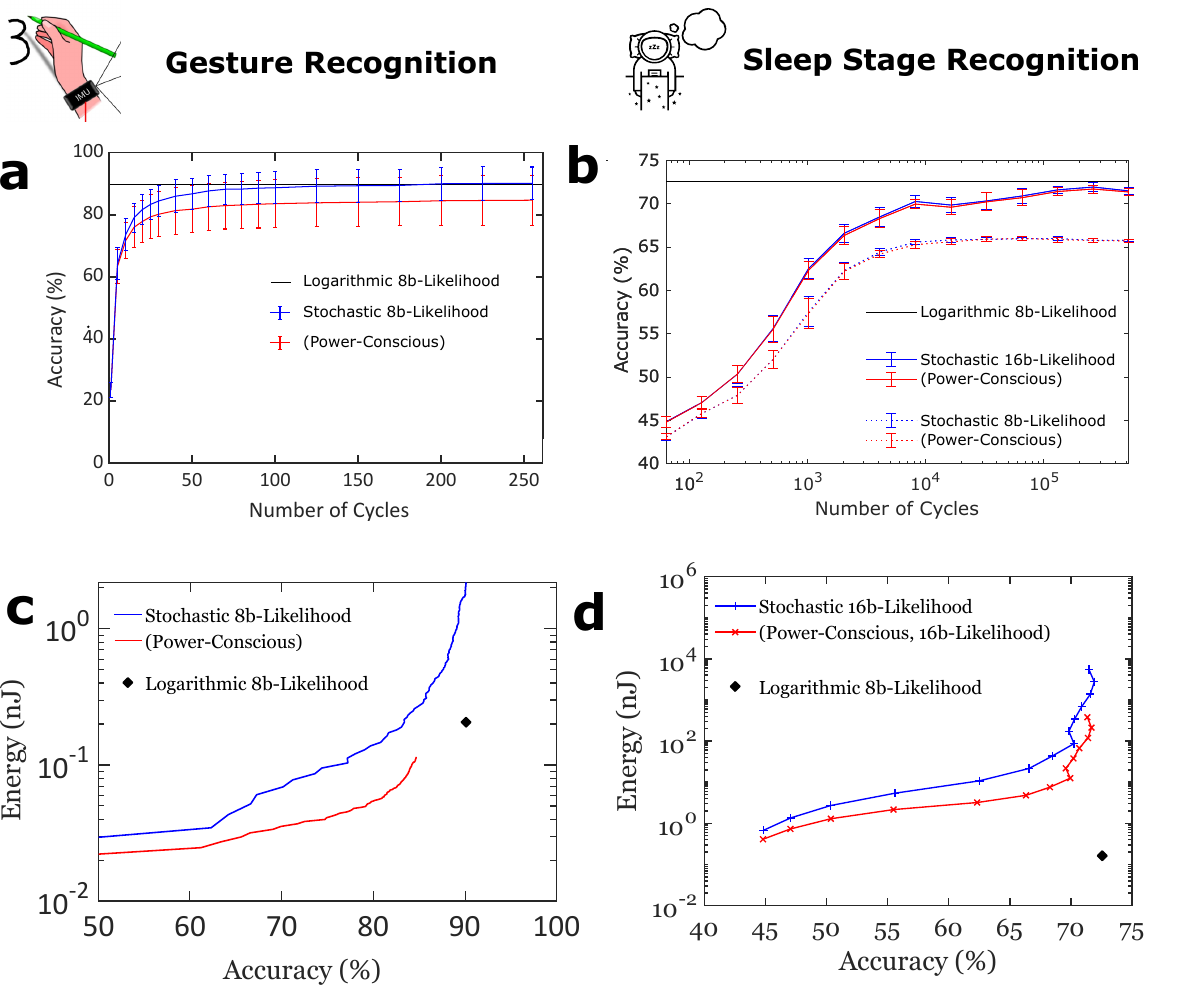}
\caption{
\textbf{Benchmarking the accuracy and the energy consumption of the stochastic and logarithmic Bayesian machines.}
\textbf{a,b} Accuracy of the  stochastic machine on \textbf{a} gesture recognition and \textbf{b} sleep stage classification, as a function of number of clock cycles, using the conventional stochastic computing or the power conscious approach. The accuracy of the logarithmic machine (one clock cycle) is plotted as a reference.  Error bars are defined in the Methods section.
\textbf{c,d} Energy consumption as a function of accuracy  for \textbf{c} gesture recognition and \textbf{d} sleep stage classification, using all approaches considered in \textbf{a,b}. This Figure is obtained associating several simulation methodologies (see Methods).
}
\label{fig:accuracy}
\end{figure}

\subsection*{Comparison of logarithmic and stochastic Bayesian machines}

We now present a comparative analysis of the stochastic and logarithmic Bayesian machines, utilizing two key tasks as benchmarks: gesture recognition,  previously discussed in ref.~\cite{harabi2023memristor}, and the sleep stage classification throughout the night. 
A specificity of stochastic computing is its inherently averaged output over multiple clock cycles. As such, an increase in the number of clock cycles directly enhances the accuracy of the results. However, this accuracy increase comes at the expense of longer latency and increased energy consumption. Fig.~\ref{fig:accuracy}a visualizes the accuracy for gesture recognition against the number of clock cycles. The logarithmic Bayesian machine, equipped with 8-bit log-likelihoods, is set as a benchmark. The presented stochastic results, extracted from \cite{harabi2023memristor}, are simulated outcomes from a scaled-up Bayesian machine in relation to the fabricated version, achieved using cycle-accurate simulations (see Methods). Two distinct strategies of stochastic computing are juxtaposed: the conventional approach, which averages outcomes over cycles, and the "power-conscious" strategy, which ceases computation once an output yields a value of one. The power-conscious strategy strongly reduces the number of cycles, but reduces accuracy, as it does not rely on statistics and assumes that the first output to produce a one is the one associated with the highest probability, which is not systemically the case.

From the data displayed in Fig.~\ref{fig:accuracy}a, we see that stochastic computing, even with just 50 cycles, delivers quality results: 82\% accuracy for the conventional strategy and 80\% for the power-conscious approach. Elevating the cycle count to 255 boosts these figures to 90\% and 84\%, respectively, while the logarithmic machine consistently achieves a 90\% accuracy.

\begin{figure}[h!]
\centering
\includegraphics[width=\linewidth]{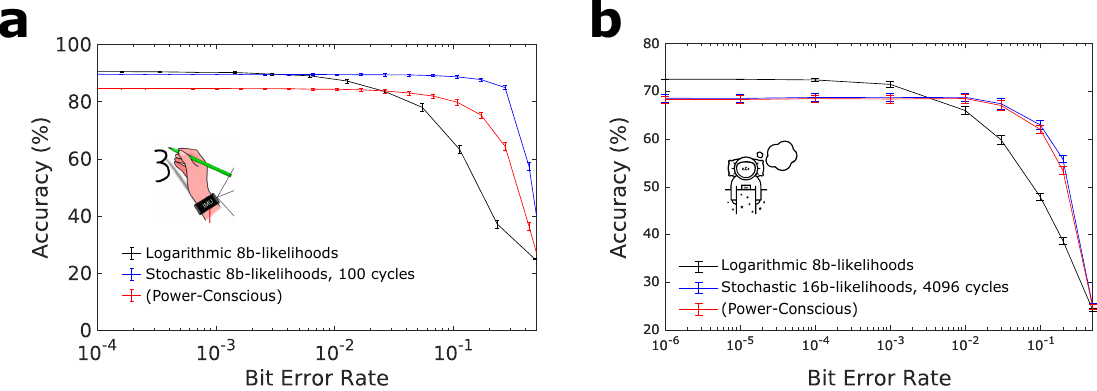}
\caption{
\textbf{Resilience of the stochastic and logarithmic Bayesian machine to memristor bit error rate.}
Accuracy of the stochastic (using conventional and power-conscious computation) and logarithmic Bayesian machines as a function of the memristor bit error rates, for \textbf{a} gesture recognition and \textbf{b} sleep stage classification tasks.
This Figure is obtained using Monte Carlo simulation (see Methods). Error bars/shadows represent one standard deviation when repeating the simulation with different memory errors. Stochastic computing simulations in  \textbf{a}  use 100 cycles, and in  \textbf{b}  4096 cycles.
}
\label{fig:robustness}
\end{figure}

When transitioning to the sleep stage classification task, the challenges posed to stochastic computing become more pronounced. As Fig.~\ref{fig:accuracy}b demonstrates, 8-bit likelihoods-based stochastic computing yields a maximum accuracy of 65\%, a stark reduction with regards to the 73\% achieved by the logarithmic approach. Switching to 16-bit likelihoods in stochastic computing does enhance its accuracy to 72\%, but requiring a staggering number of clock cycles: 10,000 cycles merely achieve 70\% accuracy. Notably, unlike the gesture recognition task, the power-conscious approach closely matches the conventional stochastic computing in performance. Note that 
the experimental measurements of Fig.~\ref{fig:sleep_stage}c use only the last 1000 points of the test dataset, while the simulation studies of Figs.~\ref{fig:accuracy} and~\ref{fig:robustness} use the complete test dataset,
explaining why the baseline differs slightly from the experiments (see Methods).

The rationale behind these outcomes is intuitive: the sleep stage task inherently deals with low likelihood values, a consequence of its time-dependent characteristics. Specifically, when in a given sleep stage, the likelihood of transitioning to another stage in the succeeding time step, denoted as 
$p\left(Y(t) = y_2 \mid Y(t-1) = y_1\right)$,
is limited. However, robust EEG and EMG data can counteract these probabilities. This scenario inherently demands the management of low probabilities, a situation where stochastic computing struggles, while logarithmic computing excels. Therefore, the difficulty of managing low probabilities in complex scenarios elucidates why 8-bit likelihoods stochastic computing falls short for this task. Furthermore, this complexity explains the high number of clock cycles required by stochastic computing, as well as the comparable performance of power-conscious and conventional stochastic computing.

We now extend our comparative analysis to the energy consumption of the machines. To quantify this metric, we employ the evaluation methodology detailed in ref.~\cite{harabi2023memristor}, which utilizes industry-standard tools for integrated circuit power assessment, along with value change dump files that represent real tasks (see Methods). These assessments are based on the 130-nanometer CMOS process and hafnium oxide memristor technology employed in our test chip.

Fig.~\ref{fig:accuracy}c plots the energy consumption per gesture recognition task as a function of accuracy for both stochastic computing strategies -- conventional and power-conscious -- and the 8-bit likelihood logarithmic machine. The relative energy efficiency of stochastic and logarithmic computation is not obvious in general: a logarithmic consumption necessitates a single clock cycle, but with arithmetic operations (integer addition) that consume much more energy than the simple AND gates used in stochastic computing.
The graph reveals that the power-conscious approach consistently outperforms the logarithmic machine in terms of energy efficiency. However, it has an accuracy ceiling of 84\%, falling short of the 90\% accuracy reached by the logarithmic method. On the other hand, conventional stochastic computing only gains an energy advantage for accuracies below 80\%, rendering it non-competitive with the logarithmic machine.

The results related to the sleep stage classification task, presented in Fig.~\ref{fig:accuracy}d, indicate a starkly different situation. Here, the logarithmic approach outclasses both stochastic strategies across all accuracy levels. Although the power-conscious method shows significant energy-saving benefits over its conventional counterpart, it remains largely less competitive than the logarithmic machine. These findings further accentuate the logarithmic machine's adaptability to time-dependent tasks involving small probability values.

Beyond performance and energy efficiency, robustness to memory read errors is an important challenge in memristor-based Bayesian machines. Fig.~\ref{fig:robustness} presents Monte Carlo simulation results, incorporating artificially induced errors in the memristor arrays (see Methods). Across both tasks, all strategies display considerable resilience to memory bit error rates, a feature attributable to the very nature of machine learning tasks. Such tasks are generally robust to errors in parameters due to inherent redundancies—for instance, multiple observations can provide overlapping information.

Interestingly, we see in Fig.~\ref{fig:robustness} that the stochastic approaches exhibit heightened resilience with regards to the logarithmic one, but not for the reasons one might initially surmise. Since memory is read once per input presentation, any error is not smoothed out by the stochastic computing; instead, it impacts the entire calculation. The greater robustness of the stochastic machines is rather attributable to the linear representation of likelihoods, where bit-flip errors are, in average, less consequential than in a logarithmic representation.

\FloatBarrier

\section*{Discussion}

Our results demonstrate that the logarithmic Bayesian machine offers a compelling advantage with regard to stochastic computing, to perform Bayesian inference in scenarios that require handling low-probability values efficiently, such as in the time-dependent task of sleep stage classification. For this task, we saw that logarithmic computation largely beats stochastic computation in terms of latency and energy consumption. Traditional stochastic Bayesian machines have their merits in simpler probabilistic calculations where high accuracy is less critical. We saw that a stochastic machine consumes less energy than the logarithmic design for gesture recognition, when accuracy lower than 84\% is targeted. 

Stochastic machines are a natural match in a near-memory computing concept, as they use extremely simple arithmetic circuits and involve reduced data movement. We also saw that they are also more robust with regards to memory bit errors than the logarithmic design. Still, stochastic machines struggle with precision and latency in  cases with low probabilities. The logarithmic approach reduces computational complexity while maintaining accuracy with regards to traditional Bayesian inference by transforming multiplicative operations into simpler integer addition. Therefore, even if it does not match the concept as naturally as the stochastic machine, is stills allows using near-memory computation, and provides an excellent compromise between the conceptual beauty of stochastic Bayesian machine and non near-memory-computing approaches. 

The logarithmic memristor-based Bayesian machine shows a robust performance across power supply conditions, without the need of any calibration, which we validated experimentally on our prototype integrated circuit. With its proficiency in efficiently handling Bayesian inferences, this machine is particularly suited for edge AI applications. These include, but are not limited to, healthcare monitoring systems where real-time decision-making based on complex data patterns is crucial. The technology's adaptability to low-power environments and its ability to provide explainable AI outputs align well with the growing regulatory and user demands for transparency in AI operations.

Future research will focus on scaling the logarithmic Bayesian machine for broader applications and further reducing its power consumption. Integration with other forms of emerging non-volatile memory technologies could also be explored to enhance performance and durability. Moreover, the adaptability of this approach to other types of probabilistic models offers a rich field for further exploration.

\section*{Acknowledgements}
This work was supported by the European Research Council starting grant NANOINFER (reference: 715872). It also benefits from France 2030 government grants managed by the French National Research Agency (ANR-22-PEEL-0010 and ANR-22-PEEL-0013) and  the support of the cleanroom RENATECH network.
The authors would like to thank  M.~Faix, R.~Frisch, E.~Mazer, A.~Renaudineau, and J.~Simatic  for discussion and invaluable feedback. 
Parts of this manuscript were revised with the assistance of a large language model (OpenAI ChatGPT).

\section*{Author contributions statement}
C.T. and K.E.H and T.H. designed the test chip, under the supervision of J.M.P and D.Q. J.M.P. designed the mixed-signal circuits of the test chip. C.T., A.P, and T.B.  performed the electrical characterization of the system. A.P. developed the sleep stage classification application. R.L. developed the gesture recognition application. C.T. and D.Q. adapted  the applications to the memristor-based Bayesian machine. J.D. and P.B developed the initial theory of the Bayesian machine. E.V. led the fabrication of the test chip. D.Q. supervised the work and wrote the initial version of the manuscript.
All authors discussed the results and reviewed the manuscript. 

\section*{Competing interests}
The authors declare no competing  interests.

\section*{Data availability}
The data measured in this study are available from the corresponding author upon  request.
The sleep stage dataset is publicly available \cite{devuyst2005dreams}.

\section*{Code availability}
The software programs used  for modeling the Bayesian machine are available from the corresponding author upon  request.


\section*{Methods}

\subsection*{Fabrication of the system}

The logarithmic Bayesian machine was fabricated using the same flow as the integrated circuits in Refs.~\cite{harabi2023memristor,jebali2024powering,bonnet2023bringing}. The CMOS part  was manufactured by a commercial provider using a low-power 130-nanometer process that incorporates four metal layers. The memristor arrays consist of a titanium nitride (TiN)/hafnium oxide (HfO$_x$)/titanium (Ti)/TiN stack. The hafnium oxide layer is 10 nanometers thick and deposited via atomic layer deposition, with the titanium layer matching this thickness. Memristors have a 300-nanometer diameter. A fifth metal layer covers the memristor arrays, which are aligned above exposed vias. The input/output pads, designed for a custom probe card, line up along one edge for easier characterization.

\subsection*{Design of the demonstrator}

Our system involves 16 memristor arrays embedded directly within the computational architecture. 
The design is similar to the one of the stochastic Bayesian machine \cite{harabi2023memristor}: we reused all memory and mixed-signal circuits, but developed a new, more automated design flow using Cadence Innovus that allowed fully-automated place-and-route.

The memristor memory arrays are structured as 64-bit cells using a two-transistor, two-memristor (2T2R) configuration, similar to  \cite{harabi2023memristor,jebali2024powering}. This setup includes thick-oxide transistors that withstand up to five volts, necessary for memristor programming and forming. An important design challenge is the need for three distinct supply voltages:
\begin{itemize}
    \item VDD (1.2 volts nominal) powers the logic and sensing circuits.
    \item VDDR (up to 5 volts) drives the word lines connected to the gate terminals of selection transistors.
    \item VDDC (up to 5 volts) powers the bit lines or source lines of the memory arrays.
\end{itemize}
Level shifters, located on each row and column of memory arrays, allow converting VDD input signals into VDDR or VDDC voltages. These level shifters use thick-oxide transistors and described in detail in the Supplementary Notes of ref.~\cite{harabi2023memristor}. The write circuits and programming strategy of the memristors are identical to the stochastic Bayesian machine and are also described in the Supplementary Notes of ref.~\cite{harabi2023memristor}.
Sensing is performed by precharge sense amplifiers \cite{zhao2009high,zhao2014synchronous,harabi2023memristor}, designed with high-threshold thin-oxide transistors to reduce energy consumption during operations.
Design and routing of the memristor arrays and their associated mixed-signal circuitry were executed manually using Cadence Virtuoso, and they were simulated with Siemens Eldo. 

Digital circuits, such as decoders, adders, and  registers, were described in SystemVerilog and synthesized using Cadence Genus and use high-threshol thin-oxide transistors. Placement and routing of the complete design were fully automated using a homemade Tool Command Language (TCL) flow in Cadence Innovus, following a hand-designed floorplan. Physical verifications, including design rule checks, layout versus schematic checks, and antenna effect checks, were conducted using Calibre EDA tools to ensure the integrity and functionality of the final design.

\subsection*{Measurements of the system}

Our test procedure is similar to the one used for the stochastic Bayesian machine \cite{harabi2023memristor}. Our system is evaluated using a custom-designed 25-pads probe card, which interfaces with a dedicated printed circuit board (PCB) connected via SubMiniature A (SMA) connectors. This PCB links the inputs and outputs of our test chip to an ST Microelectronics STM32F746ZGT6 microcontroller unit, a Keithley triple channel 2230G-30-1 power supply (providing VDD, VDDC, and VDDR), two Keysight B1530A waveform generator/fast measurement units, and a Keysight MSOS204A oscilloscope. The microcontroller is interfaced with a computer through a serial connection, and all other equipments are connected to the computer using a National Instruments GPIB connection. Testing procedures are controlled using a Python script within a Jupyter notebook, allowing a highly automated testing process.

Before deploying the Bayesian inference capabilities, each memristor undergoes a critical forming operation to establish conductive filaments. This process involves setting VDDC and VDDR to 3.0 volts each, with VDD at 1.2 volts. Each memristor is individually addressed by our chip’s digital circuitry, which applies a one-microsecond programming pulse during this operation (see Supplementary Notes of ref.~\cite{harabi2023memristor}).

Subsequent to forming, memristors are programmed to either a low-resistance state (LRS) or a high-resistance state (HRS). For LRS programming (SET mode), VDDC is raised to 3.5 volts and VDDR to 3.0 volts. Conversely, for HRS programming (RESET mode), VDDC is increased to 4.5 volts and VDDR to 4.9 volts. Programming pulses are applied in opposite polarity to each memristor depending on the desired resistance state  (see Supplementary Notes of ref.~\cite{harabi2023memristor}).

Our memristor configuration employs a two-transistor, two-memristor (2T2R) structure, where data storage is managed through complementary programming: a zero is stored by setting the left memristor (along BL) to HRS and the right (along BLb) also to HRS, while a one is stored by setting the left to LRS and the right to HRS. This complementary scheme enhances the reliability of data storage \cite{hirtzlin2020digital}. The log-likelihoods used for Bayesian inference are stored as eight-bit integers, as described in the main text.

Once programmed, the chip no longer requires high voltages. VDD, VDDR, and VVDC are all set to the same supply voltage: by default 1.2 volts,  and it can be reduced  to 0.5 volts for power conservation during extended operations. The output from the Bayesian inference processes is captured by the microcontroller and relayed back to the controlling computer for analysis and result compilation.

\subsection*{Sleep stage recognition task}

We evaluated the performance of the machine using a task focused on sleep stage classification. This task was selected for its representativeness  of  on-edge computing scenarios, where the energy supply is constrained. Sleep stage classification can provide valuable insights into sleep disorders and overall sleep quality. We employed the DREAMS Subjects Database, consisting of 20 whole-night polysomnography (PSG) recordings from healthy subjects annotated with sleep stages \cite{devuyst2005dreams}.

Given the four-input capacity of our logarithmic Bayesian machine test chip, we had to be selective in choosing the types of data to analyze. We opted for the electroencephalography (EEG) and electromyography (EMG) signals, along with prior knowledge on sleep stages. EEG is pivotal for identifying unique brain wave patterns characteristic of different sleep stages, while EMG measures muscle tone, critical for distinguishing rapid eye movement (REM) sleep from other stages.

In the dataset, sleep stages were annotated based on the Rechtschaffen and Kales criteria. We simplified that data into four categories to fit the machine's capacity of four outputs:
awake state,    REM sleep,    light non-REM sleep (combining American Academy of Sleep Medicine, or AASM,  stages 1 and 2), deep non-REM sleep (AASM stage 3).

 For model training and testing, we selected the first healthy subject in the DREAMS database. A set of 40 five-second samples for each sleep stage was used for training, while the remaining data served as test dataset.  The restricted training dataset was intentionally chosen to demonstrate the  efficacy  in a low-data context, which is a distinctive feature of Bayesian approaches \cite{bessiere2013bayesian,harabi2023memristor}.  The experimental measurements of Fig.~\ref{fig:sleep_stage}c use only the last 1000 points of the test dataset, while the simulation studies of Figs.~\ref{fig:accuracy} and~\ref{fig:robustness} use the complete test dataset.

The design of our Bayesian model was guided by medical knowledge of sleep stages, which have distinctive features in EEG and EMG signals:
\begin{itemize}
    \item Awake: presence of alpha waves in EEG (8-13Hz) and high EMG tone;
    \item REM sleep: Low EMG tone;
    \item Light  non-REM sleep: presence of alpha (8-13Hz) and theta (4-8Hz) waves in EEG; 
    \item Deep non-REM sleep: presence of delta waves (0.5-4Hz) in EEG. 
\end{itemize}

Based on this knowledge, and training set accuracy optimization, we chose three observables. Each five-second sample underwent Fast Fourier Transform (FFT) for EEG signal analysis, and power spectral density at 1.5~Hz and 9.35~Hz frequencies were designated as the first two observables. The full power of the EMG signal was also computed and used as third observable. Then the likelihoods of the Bayesian models for each stage were estimated from the training data, and modeled using a  log-normal distribution. The model was then discretized and quantized. Prior probabilities $p \left( Y(t)\mid Y(t-1)\right)$ were adjusted based on previous sleep stages to improve the model's performance. In Fig.~\ref{fig:accuracy}b, the error bars represent one standard deviation of the mean accuracy of the Bayesian machines when repeating inference ten times.

\subsection*{Gesture recognition task}

The gesture recognition task is taken from ref.~\cite{harabi2023memristor}. It is conducted using a dataset collected in our laboratory involving ten subjects. Each subject performed four distinct gestures—writing the digits one, two, three, and a personal signature in the air. The instructions were deliberately vague to encourage natural variation in gesture execution, enhancing the dataset's diversity. Gestures were captured using the three-axis accelerometer of a standard inertial measurement unit, with each gesture repeated 25 to 27 times by each subject. The recording duration varied from 1.3 to 3 seconds depending on the subject and the specific gesture. From these recordings, we extracted ten features, encompassing mean acceleration, maximum acceleration across three axes, variance of the acceleration, and both the mean and maximum jerk (rate of change of acceleration). Then we selected six most useful features (see Supplementary Notes of ref.~\cite{harabi2023memristor}).

For training the logarithmic Bayesian machine, we utilized 20 of the 25-27 recordings for each subject, reserving the remaining 5-7 recordings for testing. All reported results use cross-validation with different test/train dataset splits. The training process involved fitting Gaussian distributions to each feature's data, which was then logarithmically transformed,  discretized, and quantized to eight bits to  suit the logarithmic computing framework. For the stochastic machine, we reproduce the results obtained in ref.~\cite{harabi2023memristor}. In Fig.~\ref{fig:accuracy}a, the error bars represent one standard deviation of the mean accuracy of the Bayesian machines associated with the ten patients.

\subsection*{Simulation of scaled Bayesian machines}

The gesture recognition task requires six inputs with 64 values and, therefore, does not fit on the fabricated Bayesian machines. To evaluate this task, we simulate a scaled-up version of the Bayesian machines, introduced in ref.~\cite{harabi2023memristor} for the stochastic case, and which we adapted here to logarithmic computing. 
The scaled-up machines comprise six columns and four rows of likelihood blocks.
These arrays utilize a differential structure similar to our test chip, effectively implementing four kilobits per array.
We developed a comprehensive behavioral model of the machines using MathWorks MATLAB to simulate their functional behavior. Additionally, synthesizable SystemVerilog descriptions of the machine were created to allow for digital synthesis and hardware implementation. Test benches were established for both the MATLAB and SystemVerilog models, using consistent input files to ensure thorough testing across all potential inputs and operational cycles. Both models were meticulously verified to ensure they were equivalent under all tested conditions, confirming the accuracy and reliability of our modeling approach. SystemVerilog description of the machines were synthesized and then placed and routed using our chosen semiconductor technology. We conducted post-place-and-route simulations to evaluate the actual performance of the hardware implementation. The results from these simulations were compared with the original MATLAB model. The hardware implementation behaved as expected without any deviations.

\subsection*{Energy consumption estimates}

Our energy consumption estimates are focused on the inference phase. They are obtained using  Cadence simulation tools, as in our experiments, energy consumption is dominated by the high capacitance of the outputs in our probe tested design. We use the energy estimation methodology developed in ref.~\cite{harabi2023memristor}. 
The energy consumption attributed specifically to the memristor arrays was calculated using Siemens Eldo circuit simulations.
The broader system energy consumption was assessed using the Cadence Voltus power integrity solution framework. This part of the estimation utilized value change dump (VCD) files generated from our test benches. These files help in making the energy consumption estimates representative of real-world usage by reflecting actual device operation during inference.
One significant challenge in these estimations is the accurate modeling of the memristor arrays, which are custom components not included in standard libraries provided by foundries. To address this, we developed a custom solution using MATLAB: a compiler that transforms memristor array data into a format interpretable by Cadence Voltus (liberty file). This compiler translates the programmed likelihoods within each memory block into a liberty file that describes the functionality of the array\cite{harabi2023memristor}.

\bibliography{sample}

\end{document}